\documentclass{article}

\usepackage{arxiv}
\usepackage{graphicx}
\usepackage[utf8]{inputenc} 
\usepackage[T1]{fontenc}    
\usepackage{hyperref}       
\usepackage{url}            
\usepackage{booktabs}       
\usepackage{amsfonts}       
\usepackage{nicefrac}       
\usepackage{microtype}      
\usepackage{lipsum}
\usepackage{braket}

\title{Order parameter conditions from mutual information and symmetry conditions}

\author{
  Ivan Arraut\\
Institute of Science and Environment and FBL,\\  
 University of Saint Joseph\\
  Estrada Marginal da Ilha Verde, 14-17, Macao, China\\
  \texttt{ivan.arraut@usj.edu.mo} \\
   \And
 Wing Chi Yu \\
 Department of Physics,\\
 City University of Hong Kong,\\
 Kowloon, Hong Kong. \\
  \texttt{wingcyu@cityu.edu.hk} \\
}

\begin{document}
\maketitle

\begin{abstract}
The mutual information method has demonstrated to be very useful for deriving the potential order parameter of a system. Although the method suggests some constraints which help to define this quantity, there is still some freedom in the definition. The method then results inefficient for cases where we have order parameters with a large number of constants in the expansion, which happens when we have many degenerate vacuums. Here we introduce some additional constraints based on the existence of broken symmetries, which help us to reduce the arbitrariness in the definitions of the order parameter in the proposed mutual information method. 
\end{abstract}


\section{Introduction}

When the symmetry of a system is spontaneously broken, a vacuum degeneracy emerges naturally \cite{1}. This is a consequence of the fact that the vacuum itself does not respect some of the symmetries of the Lagrangian (Hamiltonian). These are called broken symmetries and their generators are called broken generators. There are many systems in the nature able to develop this condition called spontaneous symmetry breaking \cite{2,21}. The phenomena has several applications in condensed matter systems as well as in high-energy physics. Tied to this phenomena, is the Nambu-Goldstone theorem, which establishes that the number of broken symmetries are equal to the number of gapless particles appearing in the system and called Nambu-Goldstone bosons \cite{1, 2}. When the system experience spontaneous symmetry breaking, an order parameter appears. The order parameter is just a field showing some long correlation in the system. One way to find whether a system is able to develop long range order or not, is by using the concept of mutual information \cite{31, 32, 3,34,35}. Once it is established that the system develops long range order, then an order parameter field should be associated to this phenomena. This field can be taken as an expansion of $n$-terms with $n$ arbitrary constants. The constants are partially fixed by the trace-less conditions of the order parameter with respect to the degenerate vacuum \cite{3}. This constraint is good for the cases where there are only a few constants to fix in the expansion. However, for the cases where we have several constants, the trace-less condition alone is of limited scope. In \cite{3}, besides the traceless condition, an arbitrarily imposed normalization's condition over the constants is assumed. Although in \cite{3} it is assumed no-knowledge about the symmetries in the system, we can still assume their existence. It is known that the combination of the order-parameter plus broken generators, can create some additional constraints over the order-parameters of the system \cite{Tr,Tr1,Tr2}. In this paper, by assuming the general existence of broken symmetries, we add two important (and additional) constraints. The first one corresponds to the fact that for every $m$-set of vacuums $\ket{0}_{m}$ connected by a set of broken symmetries $Q_m$, we have the condition $\ket{0}=e^{-i\omega}\ket{0}_{2}=e^{-2i\omega}\ket{0}_{3}=...=e^{-i(m-1)\omega}\ket{0}_{m}$, with $\omega$ defining the phase difference between the degenerate vacuums. This means that each vacuum differs at most by a phase. The phase difference between neighbors vacuums will be defined in this paper. The second condition corresponds to the cyclic property of vacuums connected by broken symmetries, which suggests that the successive action of a broken generator over a series of vacuums, connected each other through the action of the same generator, will bring back the initial vacuum after $n$ successive applications of the broken generator. These two conditions complement correctly the mutual information method and they are enough for determining the form of the order parameter and they automatically include the traceless condition. The paper is organized as follows: In Sec. (\ref{s1}), we analyze the relation between mutual information and spontaneous symmetry breaking; in addition we analyze the previously proposed constraints which partially fix the order parameter of a system. In Sec. (\ref{s2}), we introduce one extra-constraint related to the existence of broken symmetries and we demonstrate that the traceless condition is just a special case of the constraints. In Sec. (\ref{appli}), we verify a few systems where we are able to predict the form of the order parameter by using the method proposed in this paper. Finally, in Sec. (\ref{conc}), we conclude.

\section{Vacuum degeneracy and spontaneous symmetry breaking}

\begin{figure}
	\centering
\includegraphics[width=0.6\textwidth]{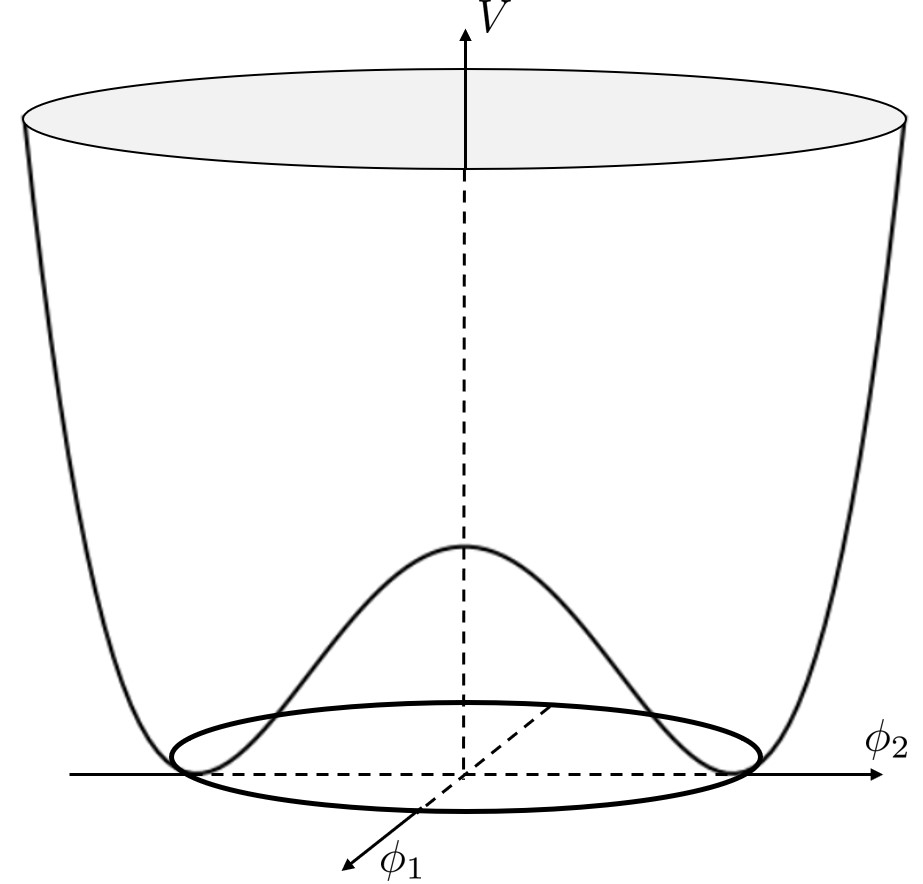}
	\caption{Typical potential of the $\sigma$-model. For certain combination of parameters, the potential develops a multiplicity of ground states (vacuums).}
	\label{Fig.1}
\end{figure}

For understanding the notions of vacuum degeneracy and spontaneous symmetry breaking, we can use as a starting point the $\sigma$-model, which contains all the necessary ingredients for understanding these concepts \cite{sigmamodel}. The model has N scalar fields defined by $\phi^i(x)$ and we can then define its Lagrangian as

\begin{equation}   \label{lagsigma}
\pounds=\frac{1}{2}(\partial_\mu\phi^i)^2+\frac{1}{2}\mu^2(\phi^i)^2-\frac{\lambda}{4}[(\phi^i)^2]^2.
\end{equation}
This Lagrangian is symmetric with respect to the transformations

\begin{equation}
\phi^i\to R^{ij}\phi^j.    
\end{equation}
This is a representation of the group $O(N)$ of orthogonal matrices in $N$-dimensions. For understanding the dynamics of the system, we can focus on its potential term, which from eq. (\ref{lagsigma}) is defined as

\begin{equation}   \label{Thispot}
V(\phi^i)=-\frac{1}{2}\mu^2(\phi^i)^2+\frac{\lambda}{4}[(\phi^i)^2]^2.    
\end{equation}
The minimum value for this potential can be obtained from the condition $\partial V(\phi^i)/\partial\phi^i=0$. The non-trivial ground state is given by 

\begin{equation}   \label{GS}
(\phi^i_0)^2=\frac{\mu^2}{\lambda}.    
\end{equation}
It is evident that this result gives us the magnitude of $\phi^i_0$. Its direction is still arbitrary but the system selects one among all the possible directions. Then for example, the system could select the ground state as follows

\begin{equation}
\phi^i_0=(0, 0, ..., v),    
\end{equation}
with $v=\mu/\sqrt{\lambda}$. This is a basic example of spontaneous symmetry breaking, where the system selects arbitrarily a vacuum state among the infinite possibilities it has. The figure (\ref{Fig.1}) illustrates the form of the potential (\ref{Thispot}) for the ground state defined in eq. (\ref{GS}). In what follows we will show an example of this phenomena, applied to ferromagnetism. 

\subsection{Example: Ferromagnetism}

One of the best examples about a multiplicity of ground states with a subsequent spontaneous symmetry breaking, is the phenomena associated to ferromagnetism. In this case we can use the theory developed by Landau \cite{sigmamodel, Landau}. The general idea is that the material develops a preferred axis of magnetization and the total magnetization $M$ is then the order parameter. This phenomena can be better perceived by defining the Gibbs-free energy $G$ as a series expansion of the total magnetization for temperatures near the critical value $T\approx T_c$ (second order phase transition) \cite{sigmamodel, Landau}

\begin{equation}   \label{Gibbs}
G(M)\approx A(T)+B(T)M^2+C(T)M^4+...    
\end{equation}
The even power expansion on $M$, is due to the symmetry under the change $M\to-M$. The perturbative expansion around $M\approx0$ is possible because near the critical point $T\approx T_c$, we have $M\approx 0$. The equilibrium condition corresponds to the case where $\partial G/\partial M=0$. Considering the fact that the parameters $B(T)$ and $C(T)$ are defined as 

\begin{equation}
B(T)=b(T-T_c),\;\;\;C(T)=c,    
\end{equation}
with $b$ and $c$ being constant terms. From these results, we can see that $B(T)$ changes sign when the temperature $T$ crosses the critical point $T\approx T_c$. Considering this issue, then the ground state condition for this system is 

\begin{eqnarray}
M=0,\;\;\;\;\;T>T_C,\nonumber\\
M=\pm\left(\frac{b}{2c}(T-T_C)\right)^{1/2}.
\end{eqnarray}
This is obtained from the condition $\partial G/\partial M=0$. If we now consider a non-zero magnetic field $H$ applied over the material and if the $T<T_C$, then the vacuum state is non-unique. This is the point where the system has to select one among the many possible ground states of the system, breaking in this way the symmetry under rotations spontaneously. It is important to remark that this phenomena of spontaneous symmetry breaking, is associated to long correlation lengths inside the material arrangement. For analyzing this situation, we can take the total magnetization as 

\begin{equation}
M=\int dx^3s({\bf x}).
\end{equation}
Here ${\bf x}$ is a vector related to the coordinates inside the material. In this way, we can express the Gibbs free-energy in eq. (\ref{Gibbs}) as follows

\begin{equation}
G=\int dx^3\left(\frac{1}{2}(\nabla s({\bf x}))^2+b(T-T_c)s^2+cs^4-Hs\right).     
\end{equation}
In this case, we include the external magnetic field $H$ and we have also included the kinetic terms $\frac{1}{2}(\nabla s({\bf x}))^2$. The variational principle gives us the result

\begin{equation}   \label{Nablela}
\left(-\nabla^2+2b(T-T_C)+4cs^2({\bf x})\right)s({\bf x})=H({\bf x}).    
\end{equation}
Here we have taken the magnetic field as a function of the position ${\bf x}$. When $T>T_C$, the magnetization almost vanishes and then we can ignore the term $4cs^2({\bf x})$ in eq. (\ref{Nablela}). Under this approximation, if we take $H({\bf x})=H_0\delta({\bf x})$, then after doing a Fourier transformation over eq. (\ref{Nablela}), we get

\begin{equation}
D({\bf x})=\frac{H_0}{4\pi}\frac{1}{r}e^{-r/\zeta},
\end{equation}
with $\zeta=(2b(T-T_C))^{-1/2}$, which is precisely the correlation length. Then for example, when $T\to T_C$, we have $\zeta\to\infty$, namely, an infinite correlation between the spins inside the material. Then there is exchange of information between distant spins inside the material when the temperature reaches the critical value. This brings us to a very important concept for analyzing the phase transitions in different materials, namely, the concept of mutual information. 

\section{The mutual information method and the construction of a local order operator}   \label{s1}

The mutual information method, developed in \cite{3}
, suggests that the existence of long range order can be deduced by evaluating a quantity which depends on the von-Neumann entropy. For understanding the method, we can start by defining a Hamiltonian for a quantum many-body system as

\begin{equation}
\hat{H}=\sum_i\hat{h}_i,    
\end{equation}
where $\hat{h}_i$ corresponds to the local Hamiltonian. Consider now two blocks $i$ and $j$, separated by the distance $\vert i-j\vert$. These blocks can consist of a single site or multiple sites, depending on the minimal sized of the subsystem that captures the long-range correlation. The reduced density matrix of a block $i$ is defined as \cite{3}

\begin{equation}
\bra{\mu'}\rho_i\ket{\mu}=tr_{\notin i}(\hat{a}_{i\mu'}\rho\hat{a}^+_{i\mu}).    
\end{equation}
The trace is done over all the degrees of freedom, except those corresponding to the block $i$ itself. The density matrix is defined as usual by the expression $\rho=\ket{\Psi}\bra{\Psi}$, where $\ket{\Psi}$ is the ground state of the Hamiltonian. Additionally, $\hat{a}_i$ corresponds to the annihilation operator acting over the states $\ket{\mu}$ on the site $i$ and satisfying the commutation or anticommutation relations. Similarly, it is possible to define the reduced density matrix for two blocks as

\begin{equation}
\bra{\mu'\nu'}\rho_{i\cup j}\ket{\mu\nu}=tr_{\notin i \cup j} (\hat{a}_{i\mu'}\hat{a}_{j\nu'}\rho\hat{a}_{j\nu}^+\hat{a}^+_{i\mu}).
\end{equation}
The trace is over all the degrees of freedom except that of $i$ and $j$.
Due to the probability interpretation of the reduced density matrix's diagonal elements, it is a matter of convention to normalize the density matrices as follows 

\begin{equation}
tr(\rho_i)=tr(\rho_j)=tr(\rho_{i\cup j})=1.    
\end{equation}
The density matrices can be diagonalized in the following form

\begin{eqnarray}
\rho_i=\sum_\mu p_\mu\ket{\psi_{i\mu}}\bra{\psi_{i\mu}},\;\;\;\;\;\rho_j=\sum_\nu p_\nu\ket{\psi_{i\nu}}\bra{\psi_{i\nu}},\;\;\;\;\;\rho_{i\cup j}=\sum_{\mu\nu}q_{\mu\nu}\ket{\phi_{i\nu}}\bra{\phi_{i\nu}}.
\end{eqnarray}
Additionally, we can define the von-Neumann entropy as follows \cite{Win1}

\begin{equation}
S=-\sum_\mu p_\mu \log_2p_\mu.    
\end{equation}
This quantity measures the amount of entanglement between the block $i$ and the rest of the system. If we want to evaluate only the correlation between a pair of blocks $i$ and $j$, then we define the mutual information as

\begin{equation}   \label{mut}
S(i\vert j)=S(\rho_i)+S(\rho_j)-S(\rho_{i\cup j}).    
\end{equation}
It has been proved before that this quantity can be used for analyzing the critical phenomena \cite{3}. In fact, the non-vanishing mutual information as it is defined in eq. (\ref{mut}) and defined between two distant blocks, means that there exists a long-range correlation between the blocks, here defined correspondingly as $i$ and $j$. Here we will not demonstrate it but the details can be found in \cite{31, 32, 3}. 

\subsection{The local order operator}

It is possible to construct a local order operator from the spectra of the density matrix \cite{3}. We can then define the order parameter as

\begin{equation}
\hat{\phi}=\sum_{\langle\mu, \nu\rangle}\left(\omega_{\mu\nu}\hat{a}^+_{i\mu}\hat{a}_{i\nu}+\omega^*_{\mu\nu}\hat{a}^+_{i\nu}\hat{a}_{i\mu}\right).    
\end{equation}
Here the coefficients $\omega_{\mu\nu}$ (and its conjugate) can be either, diagonal or non-diagonal. If $\mu=\nu$, then we have diagonal coefficients and we only need the subindex $\mu$ and in addition $\omega_{\mu\nu}=\omega_{\mu\nu}^*$ for this special case. The order parameter is diagonal or non-diagonal depending on the basis where it is expanded. In particular, it depends on the relation between the basis and the definition of the different vacuums in the same basis. Independent on whether or not the selected basis makes the order parameter diagonal, the order parameter has to satisfy the traceless condition 

\begin{equation}
tr(\hat{\phi})=0=\sum_{\mu, \nu}p_{\mu\nu}\omega_{\mu\nu}=0   \end{equation}
This condition becomes simplified when the order parameter is diagonal. In such a case, the expression is simply

\begin{equation}
\sum_{\mu}\omega_\mu p_\mu=0.    
\end{equation}
Note that $\mu\leq\zeta$, where $\zeta$ is the rank of the density matrix. The traceless conditions cannot be used if the range of the matrix $\rho$ is equal to one. This condition, is able to fix some relation between the $\omega_\mu$-coefficients. If we have $n$-coefficients of this type, then the traceless condition reduces the number of independent coefficients to $n-1$.  

\section{Additional conditions: General broken symmetries of the system}   \label{s2}

The traceless condition is useful when the amount of coefficients to fix is not so large. This happens when we have only a few degenerate vacuums. However, when we have several degenerate vacuums, then we will also have several coefficients $\omega_\mu$. Here we formulate additional conditions besides the traceless one, based on the fact that the action of a broken symmetry generator is to map one vacuum into another one, taking into account that we are dealing with vacuum degeneracy.    

\subsection{Mapping one vacuum into another one with the same energy level via broken symmetry}

Consider that we can break the symmetry of the system spontaneously by selecting one among all the possible vacuums. Consider now a broken generator defined as $Q_a$. Then we can define the condition

\begin{equation}   \label{Qa}
Q_a\ket{0}=\ket{\bar{0}}.    
\end{equation}
Here $\ket{0}\neq\ket{\bar{0}}$ (inequivalent vacuums) but both vacuums are at the same energy level. In general, we can decompose the order parameter in two components as follows 

\begin{equation}   \label{Qa2}
\hat{\phi}=\hat{\phi}_1+i\hat{\phi}_2.    
\end{equation}
We can conventionally define the component $\hat{\phi}_2$ as the one with zero vacuum expectation value for some arbitrarily selected vacuum. Let's fix such a vacuum to be $\ket{0}$, and then 

\begin{equation}   \label{Qa3}
\bra{0}\hat{\phi}_1\ket{0}\neq0,\;\;\;\;\;\;\bra{0}\hat{\phi}_2\ket{0}=0.
\end{equation}
The components can be selected such that they are related to each other through the broken symmetry generator $Q_a$ as 

\begin{equation}   \label{Qa4}
[Q_a,\hat{\phi}_2]=i\hat{\phi}_1.    
\end{equation}
This result suggests another way to say that the effect of the broken symmetry is to map one vacuum into the other. If we have $n$-degenerate vacuums connected by the same broken generator, then by applying $n$-times $Q_a$ to the vacuum state, we just get the initially fixed vacuum as follows

\begin{equation}   \label{Qa5}
(Q_a)^n\ket{0}=\ket{0}.    
\end{equation}
Then in general, each vacuum is connected to each other by one expression of the form

\begin{equation}   \label{Qa6}
\ket{0}_2=e^{-i\phi}\ket{0}_1,\;\;\;\;\;\ket{0}_3=e^{-2i\phi}\ket{0}_1, \;\;\;\;\;
\ket{0}_4=e^{-3i\phi}\ket{0}_1, ....,\ket{0}_n=e^{-(n-1)i\phi}\ket{0}_1,\;\;\;\;\; \ket{0}_1=e^{-ni\phi}\ket{0}_1,
\end{equation}
where $\phi=\alpha Q_a$ represents the effect of the broken generators over the states. If we have $m$-broken generators defined as $Q_1$, $Q_2$, $Q_3$,...$Q_m$, each generator will reproduce similar mappings, as those defined in eqns. (\ref{Qa}), (\ref{Qa4}), (\ref{Qa5}) and (\ref{Qa6}). If we have the $n$ vacuums defined previously, then we can know the phase difference between vacuums in a explicit form as

\begin{equation}   \label{important}
\phi=\frac{2\pi}{n},    
\end{equation}
for each broken generator acting over a set of vacuums. Then for the special case where we only have two vacuums, then $\phi=\pi$ because $n=2$. This situation appears for example for the case of ferromagnetism \cite{3}. Note that for the general expansion of the order parameter

\begin{equation}   \label{standardexp}
\hat{\phi}_1=\sum_n\omega_n\hat{a}^+_{i n}\hat{a}_{i n},  
\end{equation}
the relation between the coefficients $\omega_n$ is determined by the condition (\ref{important}) because each value of $n$ corresponds to a different vacuum. Then, for $n=2$, since $\phi=\pi$, we have 

\begin{equation}
\ket{0}_1=-\ket{0}_2,    
\end{equation}
from the result (\ref{Qa6}). This condition applied to the standard expansion (\ref{standardexp}), means $\omega_1=-\omega_2$ with the series expansion containing only two terms. When we have more than two vacuums, the conditions can be generalized and the complex coefficients $\omega_n$ will be related to each other by a fixed phase. Note that the traceless condition defined in \cite{3}, corresponds to a special situation here. In addition, note that here the condition suggesting that the maximum value of $\omega_n$ is one, is reduced to the condition fixing only one of the magnitudes of $\omega_i$ to one, namely, $\vert\omega_i\vert=1$. Then the difference between the different coefficients will be reduced to only phase differences. In this way, then instead of the condition (\ref{Qa6}), we can use 

\begin{equation}   \label{Thisresult}
\omega_2=e^{-i\phi}\omega_1,\;\;\;\;\;\omega_3=e^{-2i\phi}\omega_1, \;\;\;\;\;\omega_4=e^{-3i\phi}\omega_1, ....,\ket{0}_n=e^{-(n-1)i\phi}\omega_1,\;\;\;\;\; \omega_1=e^{-ni\phi}\omega_1,   
\end{equation}
expressing then the fact that all the $\omega_i$ coefficients can be expressed as a common magnitude $\vert\omega_1\vert$, times a phase. 
The cyclic property given by the last term in this equation, which defines the result (\ref{important}), gives us the following conditions 

\begin{equation}
\sum_n\vert\omega_n\vert^2=n\vert\omega_1\vert^2,   
\end{equation}
for the case of multiple vacuums, connected by broken symmetries and having the same energy level. If some vacuums are connected by broken symmetries but in addition have different energy levels, chemical potentials should appear, generating then some gap for the possible Nambu-Goldstone bosons (if any) appearing in the system. We can derive another important conditions if we sum over all the coefficients $\omega_n$ related to the different definitions of vacuum as follows

\begin{equation}
\sum_n\omega_n=\vert\omega_1\vert\left(e^{-i\phi}+e^{-2i\phi}+e^{-3i\phi}+...+e^{-ni\phi}\right).    
\end{equation}
The previous expression can be simplified if we take into account that it represents a geometric series, and then 

\begin{equation}
\sum_n\omega_n=\vert\omega_1\vert\left(\frac{1-e^{-ni\phi}}{e^{i\phi}-1}\right)=\vert\omega_1\vert\left(\frac{e^{-i\phi}-e^{-(n+1)i\phi}}{1-e^{-i\phi}}\right).    
\end{equation}
If the vacuums satisfy the cyclic property, then $e^{-(n+1)i\phi}=e^{-i\phi}$; and then we can conclude that

\begin{equation}   \label{omega0}
\sum_n\omega_n=0,   
\end{equation}
which is a natural result equivalent to the traceless condition in \cite{3}. This demonstrates that the traceless condition is just a natural consequence of the fact that the vacuums are connected by some broken symmetry plus the cyclic property summarized in the definition $e^{-ni\phi}=1$. Note that once the norm of $\omega_1$ is fixed, then the norm for all the other coefficients is also fixed as far as we consider vacuums of the same energy level. Then once we know how many vacuums are at the same energy level, for example, by analyzing the reduced density matrix spectrum, all their coefficients $\omega_\mu$ will be defined. Regarding the order of magnitude of $\omega_\mu$, by fixing the magnitude of $\vert\omega_1\vert=1$, immediately fixes the magnitude of the others (not the sign) coefficients connected by a broken generator. For any number of vacuums, the phase $\phi$ can be found from eq. (\ref{important}). In this way for example, if we have three vacuums then $n=3$ and $\phi=2\pi/3$. In this way we have 

\begin{equation}
\omega_2=-\frac{1}{2}\omega_1+\frac{\sqrt{3}}{2}i\omega_1,\;\;\;\;\;\omega_3=-\frac{1}{2}\omega_1-\frac{\sqrt{3}}{2}i\omega_1.    
\end{equation}
This result satisfies evidently the result (\ref{omega0}). 

\section{Applications}   \label{appli}

Here we will analyze different cases where we can apply our method for getting in a natural way the corresponding order parameters. 

\subsection{Ferromagnetic long-range order}

For the case of ferromagnetism, we define the unbroken vacuum as

\begin{equation}
\ket{\Psi}=\frac{1}{\sqrt{2}}\left(\ket{\uparrow\uparrow...\uparrow\uparrow}+\ket{\downarrow\downarrow...\downarrow\downarrow}\right)    
\end{equation}
This situation corresponds to a spin chain, with all the atoms in the chain having a common spin either, up or down. We consider here these two possibilities as the two independent vacuums. If the symmetry of the system is spontaneously broken, the system will select any of the two possible vacuums. If we apply our formulation to this case, then $n=2$ and then we get $\phi=\pi$ in agreement with eq. (\ref{important}), and from eq. (\ref{standardexp}), we obtain

\begin{equation}   \label{standardexp2}
\hat{\phi}_1=\omega_1\hat{a}^+_{i 1}\hat{a}_{i 1}+\omega_2\hat{a}^+_{i 2}\hat{a}_{i 2},  
\end{equation}
Since $n=2$, then $\omega_1=-\omega_2$ if we use the result (\ref{Thisresult}). Since all the $\omega_n$-coefficients related to the existence of independent vacuums (but connected through a symmetry transformation) are related to each other by a phase transformation, they can be normalized to one and then we can define them unambiguosly. Then $\vert\omega_1\vert=\vert\omega_2\vert$. This case corresponds to the ferromagnetic case showed in \cite{3}. Then the method proposed here works perfectly.

\subsection{One dimensional ferromagnetic}

Here we consider the Hamiltonian

\begin{equation}
\hat{H}=\sum_{j=1}^N\left(S_j^xS_{j+1}^x+S_j^yS_{j+1}^y+S_j^zS_{j+1}^z\right).    
\end{equation}
This Hamiltonian is basically the same of the Dimer case \cite{MG} but excluding the next nearest neighbor interactions. Here $S_j^x$, $S_j^y$ and $S_j^z$ are spin-1/2 operators at the site $j$. This model is normally solved by using the Bethe-ansatz method \cite{BA}. This case has two different vacuums connected through the condition

\begin{equation}
\omega_2=\omega^*_1.    
\end{equation}
If $\omega_1=x+iy$, then $\omega_2=x-iy$ and then the two independent vacuums corresponding to the ground state are separated by an angle 

\begin{equation}   \label{Lastetas}
\tan\left(\frac{\theta}{2}\right)=\frac{y}{x}.    
\end{equation}
For this case, we identify two independent vacuums. The order parameter is then obtained from eq. (\ref{standardexp}) as

\begin{equation}   \label{mamamama}
\hat{\phi}=\omega_1\hat{a}_{i1}^+\hat{a}_{i2}+\omega_1^*\hat{a}_{i2}^+\hat{a}_{i1}.    
\end{equation}
With the definitions given for $\omega_1$, we find that the previous expression corresponds to \cite{3}

\begin{equation}
\hat{\phi}=x\hat{O}_i^x-y\hat{O}_i^y.    
\end{equation}
Here $\hat{O}_i^{x, y}$ has a correspondence with the Pauli matrices $\hat{\sigma}_{x,y}$. This means that $\hat{O}_i^{x, y}$ represent the generators of rotations in the spin space. Then immediately we can conclude that the order parameter defined in eq. (\ref{mamamama}) has two components, corresponding to the two independent vacuums, connected each other through a combination of rotation around $x$ and rotation around $y$. This conclusion can be obtained from the expression (\ref{Qa6}), which represents the connection between the number of coefficients $\omega_i$ and the number of independent vacuums. The two independent vacuums can be defined as

\begin{equation}
\ket{0}_1=e^{i\theta}\ket{0}_2,    
\end{equation}
with the angle $\theta$ defining the rotation around the $z$-axis. The same angle is defined in eq. (\ref{Lastetas}).   
\section{Conclusions}   \label{conc}

In this paper we have learned a different way to interpret and define the order parameters in systems where there is a degeneracy on the vacuum state and as a consequence, there exists the possibility of having spontaneous symmetry breaking once a specific vacuum condition is selected. This new way to visualize the problem helps us to find additional constraints for the definition of an order parameter. Our starting point is the expansion of the order parameter in terms of the particle number operator as it was suggested in \cite{3}. Here we have explained how to constraint the coefficients $\omega_i$ of the expansion based on the fact that a broken symmetry for the vacuum, maps one of the vacuums toward another one. The order parameter itself can be expanded in a base with vectors defining the different vacuums. It is for this reason that the vacuum expectation value of an order parameter does not vanish when it is defined for a single vacuum. Other results, like the traceless condition for the order parameter emerges naturally from this scenario. This condition is a natural consequence of the expansion of the order parameter in a base defining all the possible vacuums of the system.\\\\        

{\bf Acknowledgement}\\
W. C. Yu acknowledges financial support from the National Natural Science Foundation of China (Grants No. 12005179) and City University of Hong Kong (Grant No.9610438).

\bibliographystyle{unsrt}  


\end{document}